%% file: main.tex
\documentclass[submission,copyright,creativecommons]{eptcs}
\providecommand{\event}{PLACES 2015}
\usepackage{xspace}
\usepackage{dsfont}
\usepackage{eucal}

\usepackage{amsmath}
\usepackage{amsthm}
\usepackage{amssymb}
\usepackage{amsfonts}
\usepackage{mathtools}

\usepackage{listings}
\usepackage{wrapfig}
\usepackage{graphicx}
\usepackage{lstcustom} % style=eclipse
\usepackage{color}

\definecolor{light-gray}{gray}{0.85}

\newtheorem{mydef}{Definition}
\newtheorem{theorem}{Theorem}
\newtheorem{lemma}{Lemma}

\title{Behavioural types for non-uniform memory accesses}
\author{Juliana Franco and Sophia Drossopoulou
\institute{Imperial College London, United Kingdom}
\email{\{j.vicente-franco, s.drossopoulou\} @ imperial.ac.uk}}

\def\titlerunning{Behavioural types for NUMA}
\def\authorrunning{Franco \& Drossopoulou}

\include{macros}

\begin{document}
\maketitle

\begin{abstract}
$~ \hspace{0.3cm}$
Concurrent programs executing on NUMA architectures consist of concurrent 
entities (e.g. threads, actors) and data placed on different nodes. 
Execution of these concurrent entities often reads or updates states from 
remote 
nodes. The performance of such systems depends on the extent to which the 
concurrent entities can be executing in parallel, and on the amount  of the 
remote 
reads and writes. 
\\$~ \hspace{0.4cm}$
We consider an actor-based object oriented language, and propose a type 
system which
expresses the topology of the program (the placement of the actors and 
data on the 
nodes), and an effect system which characterises remote reads and writes 
(in terms of 
which node reads/writes from which other nodes). We use a variant of 
ownership types 
for the topology, and a combination of behavioural and ownership types for 
the effect system.
\end{abstract}

\section{Introduction}
\input{introduction}

\section{Syntax}
\input{syntax}

\section{Semantics}
\input{semantics}

\section{Type Checking}
\input{typing}

\section{The global behaviour}
\input{main_results}

\section{Final Remarks}
\input{conclusion}

%%%%%%%%%%%%%%%%%%%%%
\paragraph{Acknowledgements. }
This work was funded by the EU project UpScale FP7-612985
(\url{http://www.upscale-project.eu/}).

\nocite{*}
\bibliographystyle{eptcs}
\bibliography{references}

%\newpage
\appendix
\input{appendix}

\end{document}

%% file: macros.tex
\lstdefinelanguage{btop}{
  keywords={active,class,def,create,as,if,then,else,new,for,in,do,let,null,true,false}, 
  morecomment=[l]{//}, 
  morecomment=[s]{/*}{*/}, 
  morestring=[b]",
  tabsize=2,
  literate= 
    {->}{$\rightarrow$}2
    {::}{$\parallel$}2
    {<}{$\langle$}1
    {>}{$\rangle$}1
    {leftbrace}{$\{$}1	% To be used in, e.g., footnotes
    {rightbrace}{$\}$}1
    {_1}{$_\textsf{\scriptsize 1}$}1
    {_2}{$_\textsf{\scriptsize 2}$}1
    {_i}{$_\textsf{\scriptsize i}$}1
    {_n}{$_\textsf{\scriptsize n}$}1
}
\newcommand{\keyword}[1]{\textsf{\upshape\small #1}\xspace}

\newcommand{\grmeq}{::=}
\newcommand{\grmor}{\mid}
\newcommand{\sequence}[1]{\overline{#1}}
\newcommand{\sequencePlus}[1]{\overline{#1}^+}
\newcommand{\concat}{::}
\newcommand{\getInPos}[1]{\hspace{-0.1cm}\downarrow_{#1}}
\newcommand{\ands}{\ \wedge\ }
\newcommand{\ors}{\ \vee\ }

\newcommand{\pluseffect}{\circ}

\newcommand{\retk}{\keyword{return}}
\newcommand{\defk}{\keyword{def}}

\newcommand{\intk}{\keyword{int}}
\newcommand{\boolk}{\keyword{bool}}
\newcommand{\ifk}{\keyword{if}}
\newcommand{\thenk}{\keyword{then}}
\newcommand{\elsek}{\keyword{else}}
\newcommand{\npek}{\keyword{NPE}}

\newcommand{\loopk}{\keyword{Loop}}
\newcommand{\fork}{\keyword{for}}
\newcommand{\ink}{\keyword{in}}
\newcommand{\dok}{\keyword{do}}

\newcommand{\nullk}{\keyword{null}}
\newcommand{\nilk}{\keyword{nil}}
\newcommand{\truek}{\keyword{true}}
\newcommand{\falsek}{\keyword{false}}
\newcommand{\ask}{\keyword{as}}
\newcommand{\activek}{\keyword{active}}
\newcommand{\classk}{\keyword{class}}
\newcommand{\ork}{\keyword{or}}
\newcommand{\newk}{\keyword{new}}
\newcommand{\letk}{\keyword{let}}
\newcommand{\thisk}{\keyword{this}}
\newcommand{\skipk}{\keyword{skip}}

\newcommand{\readT}[2]{\keyword{rd}(#1, #2)}
\newcommand{\writeT}[2]{\keyword{wrt}(#1, #2)}
\newcommand{\messageT}[3]{\keyword{msg}(#1, #2, #3)}
\newcommand{\msgT}[3]{\keyword{msg}(#1, #2, #3)} % dynamic
\newcommand{\parallelT}[2]{[ #1 , #2 ]} % \parallel?

\newcommand{\choiceT}[2]{ \{ #1\ \ork\ #2 \} }
\newcommand{\loopT}[2]{ \loopk (#1 \colon #2) }
\newcommand{\ownerT}[2]{ #1\langle #2 \rangle }

\newcommand{\itec}[3]{\ifk\ #1\ \thenk\ #2\ \elsek\ #3}
\newcommand{\acallc}[3]{ #1 ! #2 ( #3 ) }
\newcommand{\scallc}[3]{ #1 . #2 ( #3 ) }
\newcommand{\freadc}[2]{ #1 . #2 }
\newcommand{\fwritec}[3]{ #1 . #2 = #3 }
\newcommand{\newc}[1]{ \newk\ #1 }
\newcommand{\forc}[3]{ \fork\ #1\ \ink\ #2\ \dok\ #3 }
\newcommand{\letc}[3]{ \letk\ #1 = #2\ \ink\ #3 }

\newcommand{\arraygetc}[2]{ #1 [ #2 ] }
\newcommand{\arraysetc}[3]{ #1 [ #2 ] = #3 }

\newcommand{\entname}[1]{\mathit{ #1 }}

\newcommand{\nodeconf}{\mathcal{N} }
\newcommand{\lheap}{h}

\newcommand{\stack}{\sigma}
\newcommand{\sframe}{\varphi}
\newcommand{\Aaddr}{\alpha}
\newcommand{\cloc}{\kappa}
\newcommand{\queue}{Q}
\newcommand{\context}{[ \cdot  ]}
\newcommand{\natnumber}{\mathds{N}}
\newcommand{\commk}{b}
\newcommand{\ctk}{\mathit{b}}
\newcommand{\bop}{\mathit{bop}}

\newcommand{\Gammas}{\overline\Gamma}
\newcommand{\threadk}{\mathcal{T}}
\newcommand{\thread}[2]{\langle#1, #2 \rangle}
\newcommand{\address}[2]{\langle#1 . #2\rangle}

\newcommand{\ots}{\entname{ot}}
\newcommand{\cds}{\entname{Cd}}
\newcommand{\fds}{\entname{Fd}}
\newcommand{\mds}{\entname{Md}}
\newcommand{\vars}{\entname{var}}
\newcommand{\vals}{\entname{val}}
\newcommand{\clocs}{\entname{NodeId}}

\newcommand{\locsMap}{\mathcal{L}}

\newcommand{\Nothing}{\varepsilon}
\newcommand{\Empty}{\emptyset}

\newcommand{\rulename}[1]{\textsc{\footnotesize{[#1]}}}
\newcommand{\infer}[3]{\begin{array}{c} \hfill\rulename{#1} \\ #2 \\ \hline  #3 \end{array}}
\newcommand{\inferNN}[2]{\frac{\begin{array}{c}#1\end{array}}{\begin{array}{c}#2\end{array}}} % 

\newcommand{\lexec}[3]{{#1}\overset{#3}{\rightarrow}{#2}}

\newcommand{\localrule}[3]{\infer{S#1}{#2}{#3}}

\newcommand{\gexec}[3]{{#1}\overset{#3}\rightarrow{#2}}
\newcommand{\globalrule}[3]{\infer{Gs#1}{#2}{#3}}

\newcommand{\typing}[4]{#1 \vdash #2 \triangleright #3, #4}
\newcommand{\typerule}[3]{\infer{T-#1}{#2}{#3}}
\newcommand{\typingNoC}[5]{#1, #2 \vdash #3 \triangleright #4, #5}

\newcommand{\mlookup}[2]{\mathcal{M}(#1, #2)}
\newcommand{\flookup}[2]{\mathcal{F}(#1, #2)}
\newcommand{\getownership}[2]{\operatorname{owners}(#1, #2)}
\newcommand{\update}[3]{#1[ #2 \concat #3 ]}
\newcommand{\getFieldTypes}[1]{\mathcal{F}_s(#1)}

\newcommand{\isActive}[1]{\operatorname{isActive}(#1)}
\newcommand{\isMain}[2]{\operatorname{isMain}(#1, #2)}
\newcommand{\fresh}{\operatorname{fresh\ in}}
\newcommand{\initf}[1]{\operatorname{init}(#1)}

\newcommand{\owners}[1]{\mathcal{O}(#1)}
\newcommand{\isvalid}[1]{\operatorname{isValid}(#1)}
\newcommand{\replace}[3]{#1 [ #3 ]}
\newcommand{\initialOb}[1]{\operatorname{initObj}(#1)}

\newcommand{\getHeaps}[1]{\operatorname{heaps}(#1)}
\newcommand{\typeOf}[2]{\operatorname{typeOf}(#1, #2)}
\DeclareMathOperator{\filterk}{filter}
\newcommand{\filterf}[1]{\filterk(#1)}

\newcommand{\wellFormedC}[1]{P \vdash #1}

\newcommand{\wellFormedh}[2]{#1 \vdash #2}
\newcommand{\wellFormedS}[2]{#1 \vdash #2}
\newcommand{\wellFormedConfig}[1]{\vdash #1}
\newcommand{\wellFormedNode}[2]{#1 \vdash #2}
\newcommand{\wfValue}[3]{#1 \vdash #2 : #3}

\newcommand{\getSigma}[2]{#1 \blacktriangleright #2}
\newcommand{\getSigmaAux}[2]{#1 \blacktriangleright #2}

\newcommand{\SigmaRelation}[3]{#1 \sqsubseteq_{#2} #3 }
\newcommand{\buildGammas}[2]{\operatorname{buildContext}(#1, #2)}

\newcommand{\effectRelation}[3]{#1 \sqsubseteq_{#3} #2}
\newcommand{\dom}[1]{\operatorname{dom}(#1)}

\DeclareMathOperator{\sourcek}{source}
\newcommand{\sourcef}[1]{\sourcek(#1)}
\DeclareMathOperator{\destk}{dest}
\newcommand{\destf}[1]{\destk(#1)}

%% file: introduction.tex
\newcommand{\lname}{$\mathcal{L}_{numa}$}

\begin{wrapfigure}{r}{0.49\textwidth}
\includegraphics[scale=0.6]{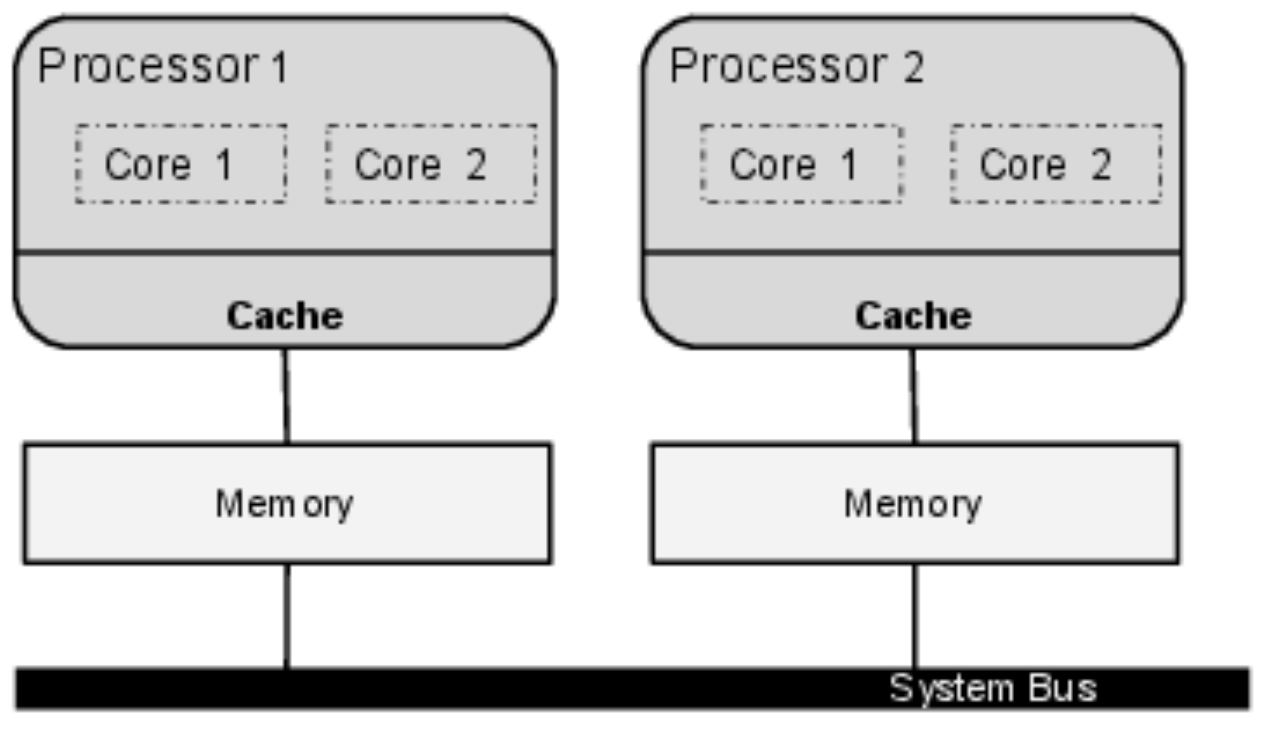}
\caption{NUMA system~\cite{NUMA}.}
\label{fig:numa_intro}
\end{wrapfigure}

A prevalent paradigm in high performance machines is NUMA (non uniform 
memory access) systems, e.g., the AMD Bulldozer server\cite{Bulldozer}. 
NUMA systems have many {\em nodes} which contain processors and memory;  
Figure~\ref{fig:numa_intro} shows the common NUMA structure. 
The nodes are connected with the other nodes through a
system bus that allows processes running on a specific node
to access the memory of the other nodes. 

Memory access is either local, i.e. accessing  
memory in the local node, or remote, i.e. accessing memory of remote nodes. 
Remote accesses require 
requests to the system bus, and are thus more expensive than local accesses. 
Moreover, different remote accesses do
not necessarily have the same cost (the time to obtain/write data in memory). 
Therefore, to characterize the communication (read/write) costs of a concurrent
 program, we need to know its topology  (the placement of the actors and data 
on the nodes), and a characterisation of the reads and writes across nodes.

In this work we consider a concurrent language based on actors (or active 
objects) and objects~\cite{Clarke:active_objects_ownership}, which we call 
\lname, a language where, for the sake of simplicity, mutually recursive 
(synchronous and asynchronous) 
method invocations with communication are assumed to be not allowed and all 
the active objects must be created in the main class. 

We develop a variant of ownership types~\cite{clarke_potter_noble:ownership_types} 
to express the location of actors and of data. In particular, we propose two 
levels of abstraction: classes have ownership (location) parameters, the main 
program defines the abstract locations and creates objects in these abstract 
locations; and at runtime the abstract locations are mapped to nodes 
(cf.\ Appendix C). We also propose a  combination of behavioural and ownership
types to characterise the interactions (reads, writes and messages sent) among 
objects located in different  nodes.

Ownership types~\cite{Clarke_oopsla12,clarke_potter_noble:ownership_types,Noble98flexiblealias} 
were first introduced to statically describe the heap 
topology. Here we introduce ownership-like annotations to describe
the system topology, that is, its nodes and where threads are running 
and data is allocated.
Behavioural 
types~\cite{contracts,honda.vasconcelos.kubo:language-primitives,Kobayashi2001,pierce_sangiorgi,typestate86} 
are usually used to describe and statically,
or dynamically, verify patterns of interaction between 
processes/threads/participants of concurrent and parallel computations.
Here we present a type system that allows the programmer to specify
the interactions among objects located in different nodes, and therefore
we abstract the communication made through the system bus.

\emph{Outline.} This paper is organised as follows: Section~\ref{sect:syntax} 
introduces the syntax of \lname, Section~\ref{sect:semantics} gives the 
operational semantics, Section~\ref{sect:typing} presents the typing rules, and 
Section~\ref{sect:typing} shows properties of \lname, and finally Section 6 
concludes. Several definitions are given in the appendix.

%% file: syntax.tex
\label{sect:syntax}

Figure~\ref{fig:syntax} presents the syntax of \lname.
A program consists of a set of class declarations representing actors and passive objects.
The use of the keyword $\keyword{active}$ in a class declaration 
indicates that the class represents actors. 
Passive objects are similar to ordinary Java objects while actors have all the properties 
of passive objects, and in addition also have their own execution thread and may send 
messages to other actors. 
As in actor-based languages, messages are stored in private queues. 
A more detailed definition can be found in~\cite{Clarke:active_objects_ownership}. 
\input{classes_syntax}

Each class, active or passive, is annotated with a set of location parameters 
$p_1, \dots ,p_n$ where $p_1$ represents the place where the instance of the class 
is allocated and $p_2, \dots,p_n$ locations that can be used in the types of the rest 
of the class. 
The location parameters of the main class, $L_1, ..., L_n$,
are abstractions of the concrete nodes, and at runtime will be mapped
to concrete node identifiers.

A class declaration might have field and method declarations. A field declaration 
consists of a field identifier and a type; 
a method declaration consists of a method identifier, one parameter 
(variable and type), 
return type, behavioural type and an expression (method body). \lname\ has the types 
$\boolk$, $\nilk$, and an ownership type $\ownerT{C}{l_1, ..., l_n}$ which represents 
objects located in $l_1$ that may contain references to objects in locations $l_2,...,l_n$. 
The syntax of expressions is similar to other OO programming languages; note only the 
asynchronous method call (message sending), $e!m(e)$.

The most interesting part of the syntax is our treatment of behavioural types. 
We have basic operations,  $\pi$,  which are reading from a remote node 
($\readT{l}{l}$), writing to a remote node ($\writeT{l}{l}$), and message 
sending ($\messageT{l}{l}{m}$)---this has to be reflected in the behaviour, as 
it adds messages to queues in remote memory. For all of them the first location
is where the expression is being executed and the second is the location where a 
read/write is made or a message sent.  
We also have types to describe conditional expressions, $\choiceT{\ctk}{\ctk}$, 
(the two branches in the if-then-else expression imply two branches in the type), 
and for-loops, $\loopT{n}{\ctk}$.  
A behavioural type, $\ctk$, may be empty, $\Nothing$, meaning that there is no
``communication'' across different nodes, the  sequence of operations, 
$\bop . \ctk$, and two types in parallel, $\parallelT{\ctk}{\ctk}$, introduced 
by message sendings.

%% file: classes_syntax.tex
\begin{figure}[!ht]
\begin{minipage}{0.5\textwidth}
% \small
\begin{align*}
P \in \entname{Program} 
  & \grmeq \cds^*  \\
\cds \in \entname{ClassDecl} 
  & \grmeq [ \activek ]\ \classk\ \ownerT{C}{\sequencePlus{p}}\ 
  \sequence{\fds}\ 
  \sequence{\mds}\\
%\entname{Main} & \grmeq 
%  \classk\ \ownerT{C}{\sequencePlus{L}}\
%  \sequence\fds\ 
%  \sequence\mds\\
%
\fds \in \entname{FieldDecl} & \grmeq f: T\\
\mds \in \entname{MethodDecl} & \grmeq \defk\ m(x: T):\ T\ \ask\ \commk\ \{ e \}\\
\ots \in \entname{OwnershipType} & \grmeq \ownerT{C}{\sequencePlus{l}} \\
T \in \entname{Type} & \grmeq \boolk 
  \grmor \nilk
  \grmor \boxed{\intk}
  \grmor \ots\\
  %\grmor \ots [n]
  % \\
%
l \in \entname{Location} & \grmeq p \grmor L \\
\vals \in \entname{Value} & \grmeq \nullk 
  \grmor \truek 
  \grmor \falsek\\
\vars \in \entname{Variable} & \grmeq x
  \grmor \thisk
\end{align*}
\end{minipage}
\begin{minipage}{0.5\textwidth}
% \small
\begin{align*}
e \in \entname{Expr} & \grmeq \vars 
  \grmor \vals
  \grmor \itec{e}{e}{e} \\ & \quad
  \grmor \scallc{e}{m}{e} 
  \grmor \acallc{e}{m}{e}
  \grmor \freadc{e}{f}  \\
  & \quad 
  \grmor \fwritec{e}{f}{e} 
  \grmor \newc{\ots}  \\
  & \quad 
  \grmor \forc{i}{n_1 .. n_2}{e} \\
  & \quad
  \grmor \letc{x}{e}{e}
  \grmor \boxed{\retk\ e}\\
\pi \in \entname{RemAccess} & \grmeq 
  \readT{l}{l}
  \grmor \writeT{l}{l}
  \grmor \messageT{l}{l}{m}\\
\bop & \grmeq
  \pi
  \grmor \choiceT{\ctk}{\ctk}
  \grmor \loopT{n}{\ctk} \\
\ctk \in \entname{Behaviour} & \grmeq
  \Nothing
  \grmor \bop. \ctk
  \grmor \parallelT{\ctk}{\ctk}
\end{align*}
\end{minipage}
\caption{Syntax of classes and (behavioural) types. 
  The boxed constructs are not user syntax.
  In the class declaration $\entname{Cd}$ the use of $[\ ]$ 
  means that the keyword $\keyword{active}$ is optional.
}
\label{fig:syntax}
\end{figure}

%% file: semantics.tex
\label{sect:semantics}

We now  describe  the dynamic semantics of \lname.
Nodes, $\mathcal{N}$, defined in Figure~\ref{fig:entities}, aim to reflect NUMA nodes.
Namely, a node in our formalism  has an identifier, a heap with all the data allocated 
in it, 
and several execution threads $\entname{EThread}$. 
An execution thread belongs to an actor, and 
has a stack and an expression being executed.
\input{entities}
A heap is a mapping from addresses to (passive and active) objects.
An object consists of a class identifier $C$, a sequence of node identifiers 
representing the 
actual  location parameters,
a mapping from field identifiers to their values, and a message queue, 
where the queue of a passive object is $\bullet$.
An address, $\alpha$, consists of a node identifier, $\cloc \in \clocs $, and an 
offset, $n\in \natnumber$.

In our system, a configuration $\sequence\nodeconf$ can be reduced to another 
configuration $\sequence\nodeconf'$ either without any communication or implying 
a remote access from one of the nodes to another node.
In the first case, the rule \rulename{GsExec1} should be applied, where only one 
node is involved in the reduction. In the second case the rule \rulename{GsExec2}
should be used, where two nodes are involved in the reduction, as shown in 
Figure~\ref{fig:global}.
%
%In both cases, a configuration is reduced when one 
%of the execution threads (its stack and its expression) reduces to another one,
\input{global_rules_fig}

In the same way, expression reduction may result in accessing remote memory or not; 
therefore we divide the operational semantics rules as follows:
\begin{enumerate}
\item Expressions that do not access memory or send messages. 
These are defined in Figure~\ref{fig:no_access}.
\item Expressions that result in accesses to memory.
These are defined in Figure~\ref{fig:local} and are further divided in:
\begin{enumerate}
\item The access happens locally---only one node required.
\item The access happens remotely---two different nodes required.
\end{enumerate}
\end{enumerate}

Figure~\ref{fig:no_access} shows shows the rules for the point 1,
where no accesses to memory, eiher in the same node or not, are made.
\input{only_local_semantics}
Each rule takes a node identifier, its heap, a stack and 
an expression, and reduces to a new heap, a new stack and a new expression.
They have the form
$\lexec{\cloc, \lheap, \stack, e}{\lheap', \stack', e'}{\pi}$.
These rules show reduction without any communication among different nodes 
(they show reduction through $\Nothing$).
The intuition behind them is standart and similar can be found in the
literature.
Note only the rule for the message receiving, $\rulename{SReceiveL}$, which takes
an empty stack and a $\nullk$ expression, meaning that the expression of the 
thread being executed is fully reduced, and returns a new frame and expression
after taking the next message in the queue to be processed.
The expression returned is the body of the method asynchronously invoked, as 
there is a new frame with the values passed as arguments.
% , which should be removed
% from the stack once the expression in fully reduced. 

Figure~\ref{fig:local} shows the semantic rules for the point 2.
The rules on the left belong to 2(a); 
they have the same form of the rules introduced in Figure~\ref{fig:no_access}.
%they take a node identifier, its heap, a stack and 
%an expression, and reduce to a new heap, a new stack and a new expression.
%They have the form
%$\lexec{\cloc, \lheap, \stack, e}{\lheap', \stack', e'}{\pi}$.
%
The rules on the right belong to 2(b); 
they take two node identifiers, their heaps, a stack and an
expression, and reduce to two new heaps, a new stack and a new expression.
They have the form
$\lexec{\cloc_1, \lheap_1, \stack, e \parallel \cloc_2, \lheap_2}
           {\lheap_1', \stack', e' \parallel \lheap'_2}{\pi}$.
In both cases they reduce through an operation described by 
$\pi$---the remote operation made or empty, $\Nothing$ (in the case of the
absence of a remote operation).
%
\input{local_semantics}
For instance, message sending in rule $\rulename{SMsgL}$
adds a message to the queue of an actor in the same
node as $\thisk$,
while  $\rulename{SMsgR}$ adds the message to the
queue of an object in a different node.
In the first case $\pi$ is empty and in the second case it is
$\msgT{\cloc_1}{\cloc_2}{m}$.
In both cases,
the stack remains unchanged and the returned expression is $\nullk$; 
namely execution is asynchronous.
All the other rules, except the context rules, on the left show,
as expected, reads and writes to the local heap and on the right
present reads and writes to a  remote heap.

%% file: entities.tex
\begin{figure}[!ht]
\begin{minipage}{0.4\textwidth}
% \small
\begin{align*}
\nodeconf \in \entname{Node} &=
  \clocs \times \entname{Heap} \times \sequence{\entname{EThread}} \\
\threadk \in \entname{EThread} &=
  \entname{Stack} \times \entname{Expr}\\
\lheap \in \entname{Heap} & =
   \entname{Addr} \rightarrow
   \entname{Object}\\
\stack \in \entname{Stack} & = \entname{Addr} \times \sequence{\entname{Frame}}\\
\sframe \in \entname{Frame} & = \vars \rightarrow \entname{value}\\
\queue \in \entname{Queue} & \grmeq\ \bullet\ \grmor\ \Empty\ \grmor m(v)\concat \queue \\
\locsMap \in \entname{LocsMap} & = \entname{LocId} \rightarrow \clocs\\
\cloc \in \clocs & =  \natnumber \\
l \in \entname{Location} & \grmeq \text{ as before } \grmor \cloc
\end{align*}
\end{minipage}
\hspace{0.15cm}
\begin{minipage}{0.5\textwidth}
% \small
\begin{align*}
o \in \entname{Object} & = \entname{ClassId}
                           \times \sequence\clocs \times \\ &
                           (\entname{FieldId} \rightarrow \entname{value})
                           \times \entname{Queue}\\
\alpha \in \entname{Addr} & = NodeId \times \natnumber\\
v \in \entname{value} & = \vals
  \grmor \entname{Addr}
  \grmor \skipk
  \grmor \npek \\
E[] & \grmeq \context\ 
  \grmor\ \scallc{\context}{m}{e}
  \ \grmor\ \scallc{\Aaddr}{m}{\context}
  \ \grmor\ \acallc{\context}{m}{e} \\ &\quad
  \ \grmor\ \acallc{\Aaddr}{m}{\context} 
  \ \grmor\ \freadc{\context}{f}
  \ \grmor\ \fwritec{\context}{f}{e} \\ &\quad
  \ \grmor\ \fwritec{\Aaddr}{f}{\context} 
  \ \grmor\ \letc{x}{\context}{e} \\& \quad
  \ \grmor\ \itec{\context}{e_1}{e_2}
  \ \grmor\ \letc{x}{\context}{e} \\ &\quad
  \ \grmor\ \retk\ \context
%%%%%%%%%%%%%%%%%%%%%%%%%
\end{align*}
\end{minipage}
\caption{Dynamic Entities. We assume the existence of a map $\locsMap$ 
that maps abstract locations (declared by the programmer in the main class) 
to NUMA node identifiers.}
\label{fig:entities}
\end{figure}

%% file: global_rules_fig.tex
\begin{figure}[!ht]
% \small
\begin{gather*}
\globalrule{Exec1}{
  \lexec{\cloc, \lheap, \stack, e}
        {\lheap', \stack', e'}
        {\pi}
}{
  \gexec{\sequence\nodeconf, 
       (\cloc, \lheap, \sequence\threadk, \thread{\stack}{e})}
      {\sequence\nodeconf, 
       (\cloc, \lheap', \sequence\threadk, \thread{\stack'}{e'})}
      {\pi} 
}
%%%%%
\\
\globalrule{Exec2}{
  \lexec{\cloc_1, \lheap_1, \stack_1, e_1 \parallel \cloc_2, \lheap_2}
        {\lheap'_1, \stack'_1, e'_1 \parallel \lheap'_2}
        {\pi}
}{
  \gexec{\sequence\nodeconf, 
      	  (\cloc_1, \lheap_1, \sequence{\threadk_1},\thread{\stack_1}{e_1}), 
      	  (\cloc_2, \lheap_2, \sequence{\threadk_2})}
        {\sequence\nodeconf, 
          (\cloc_1, \lheap'_1, \sequence{\threadk_1},\thread{\stack'_1}{e'_1}), 
          (\cloc_2, \lheap'_2, \sequence{\threadk_2})}
        {\pi} 
}
\end{gather*}
\caption{Global semantics}
\label{fig:global}
\end{figure}

%% file: only_local_semantics.tex
\begin{figure}[!ht]
% \small
\begin{gather*}
\localrule{IfTrue}{
}{
\lexec{\cloc, \lheap, \stack, \itec{\truek}{e_1}{e_2}}
      {\lheap, \stack, e_1}{\Nothing}
}
%%%
\quad
\localrule{IfFalse}{
}{
\lexec{\cloc, \lheap, \stack, \itec{\falsek}{e_1}{e_2}}
      {\lheap, \stack, e_2}{\Nothing}
}
%%%
\\
\localrule{Let}{
x \fresh \sframe
\quad
\sframe' = \sframe[x \mapsto v]
}{
\lexec{\cloc, \lheap, \stack.\sframe, \letc{x}{v}{e}}
      {\lheap, \stack.\sframe', e}{\Nothing}
}
%%%
\quad
\localrule{Ret}{
}{
  \lexec{\cloc, \lheap, \stack . \sframe, \retk\ v }
        {\lheap, \stack, v}{\Nothing}
}
%%%
\quad
\localrule{Var}{
  \sframe(x) = v
}{
  \lexec{\cloc, \lheap, \stack.\sframe, x}
        {\lheap, \stack.\sframe, v}
        {\Nothing}
}
%%%
\\
\localrule{For}{
x \fresh \sframe
\quad
e' = (\letk\ x = e \ink\ \forc{i}{(n_1+1) .. n_2}{e})
}{\lexec{\cloc, \lheap, \stack.\sframe, \forc{i}{n_1 .. n_2}{e}}
        {\lheap, \stack.\sframe[i\mapsto n_1], e'}
        {\Nothing}
}
%%%
\quad
\localrule{ForSkip}{
n_1 > n_2
}{\lexec{\cloc, \lheap, \stack, \forc{i}{n_1 .. n_2}{e}}
        {\lheap, \stack, \skipk}{\Nothing}
}
%%%
\\
\localrule{Skip}{
}{
  \lexec{\cloc, \lheap, \stack, \skipk}
        {\lheap, \stack, \nullk}
        {\Nothing}
}
\quad
\localrule{CallL}{
    \getownership{\lheap}{\alpha} = \ownerT{C}{\sequence{\cloc}}
    \quad
    \sframe = \alpha \cdot (\thisk \mapsto \alpha, x \mapsto v)
}{
  \lexec{\cloc, \lheap, \stack, \alpha.m(v)}
        {\lheap, \stack . \sframe, 
                 \retk\ 
                 \replace{\mlookup{C}{m}\getInPos{3}}{C}{\sequence{\cloc}} }  
        {\Nothing}
}
\\
\localrule{ReceiveL}
{
  \alpha\getInPos{1} = \cloc
  \quad
  \lheap(\alpha) = (C, \sequence{\cloc}, \_, m(v) \concat Q)
  \quad
  e = \replace{\mlookup{C}{m}}{C}{\sequence{\cloc}}
}
{
  \lexec{\cloc, \lheap, \alpha \cdot \Empty , \nullk }
        {\lheap[\alpha \mapsto Q], \alpha \cdot (\thisk \mapsto \alpha, x \mapsto v), \retk\ e}
  {\Nothing}
}
%%%
\\
\localrule{ContextNPE}{
}{
\lexec{\cloc, \lheap, \stack, E[\npek]}{\lheap, \stack, \npek}{\Nothing}
}
\quad
\localrule{NPE}{
}{
\lexec{\cloc, \lheap, \stack, e_{npe}}{\lheap, \stack, \npek}{\Nothing}
}\\
\text{where } e_{npe} \text{ can be } 
  \freadc{\nullk}{f},\ 
  \fwritec{\nullk}{f}{e},\ 
  \scallc{\nullk}{m}{e},\
  \acallc{\nullk}{m}{e},\ 
  \arraygetc{\nullk}{i},\ 
  \arraysetc{\nullk}{i}{e'}
\end{gather*}
\caption{Semantic rules for expressions that do not perform remote operations. 
        Null-pointer exceptions included.}
\label{fig:no_access}
\end{figure}

%% file: local_semantics.tex
%%%%%%%%%%%%%%%%%%%%%%%%%%%%%
%%%%%%%%%%%%%%%%%%%%%%%%%%%%%%%
\begin{figure}[!ht]
\begin{gather*}
\localrule{MsgL}{
\lheap' = \update{\lheap}{\address{\cloc}{n}}{m(v)}
}{
\lexec{\cloc, \lheap, \stack, \address{\cloc}{n}!m(v)}
      {\lheap', \stack, \nullk}
      {\Nothing}
}
%%%
\quad
\localrule{MsgR}{
  \begin{array}[b]{c}
    \pi = \msgT{\cloc_1}{\cloc_2}{m}
    \quad
    \lheap'_2 = \update{\lheap_2}{\address{\cloc_2}{n}}{m(v)}
  \end{array}
}{
\lexec{\cloc_1, \lheap_1, \stack, \address{\cloc_2}{n}!m(v) 
       \parallel
       \cloc_2, \lheap_2}
      {\lheap_1, \stack, \nullk \parallel \lheap'_2}
      {\pi}
}
%%%
\\
\localrule{FReadL}{
}{
\lexec{\cloc, \lheap, \stack, \address{\cloc}{n}.f}
      {\lheap, \stack, \lheap(\cloc . n)(f)}
      {\Nothing}
}
%%%
\quad
\localrule{FReadR}{
  \pi = \readT{\cloc_1}{\cloc_2}
  \quad
  v = \lheap_2(\address{\cloc_2}{n})(f)
}{
\lexec{\cloc_1, \lheap_1, \stack, \address{\cloc_2}{n}.f 
       \parallel \cloc_2, \lheap_2}
      {\lheap_1, \stack, v \parallel \lheap_2}
      {\pi}
}
%%%
\\
\localrule{FWriteL}{
}{
\lexec{\cloc, \lheap, \stack,  \address{\cloc}{n}.f = v}
      {\lheap[\address{\cloc}{n}, f \mapsto v], \stack, v}
      {\Nothing}
}
%%%
\quad
\localrule{FWriteR}{
  \pi = \writeT{\cloc_1}{\cloc_2}
  \quad
  \lheap'_2 = \lheap_2[\address{\cloc_2}{n}, f \mapsto v]
}{
\lexec{\cloc_1, \lheap_1, \stack, \address{\cloc_2}{n}.f = v
       \parallel \cloc_2, \lheap_2}
      {\lheap_1, \stack, v \parallel \lheap'_2}
      {\pi}
}
%%%
\\
\localrule{NewL}{
  \begin{array}[b]{c}
    %\ots = \ownerT{C}{L_1, \_}
    %\quad
    \cloc = \locsMap(L_1)
    \quad
    \address{\cloc}{n} \notin \dom\lheap
    \\
    \lheap' = \lheap [ \address{\cloc}{n} \mapsto \initialOb{\ownerT{C}{\sequence{L}}} ]
  \end{array}
}{
\lexec{\cloc, \lheap, \stack, \newc{\ownerT{C}{\sequence{L}}}}
      {\lheap', \stack, \address{\cloc}{n}}
      {\Nothing}
}
%%%
\quad
\localrule{NewR}{
    %\ots = \ownerT{C}{L_1, \_}
    %\quad
    \cloc_2 = \locsMap(L_1)
    \quad
    \address{\cloc_2}{n} \notin \dom{\lheap_2}
    \quad
    \pi = \writeT{\cloc_1}{\cloc_2}
    \\
    \lheap_2' = \lheap_2 [ \address{\cloc_2}{n} \mapsto \initialOb{\ownerT{C}{\sequence{L}}} ]
}{
\lexec{\cloc_1, \lheap_1, \stack, \newc{\ownerT{C}{\sequence{L}}} \parallel \cloc_2, \lheap_2}
      {\lheap_1, \stack, \address{\cloc_2}{n} \parallel \lheap_2'}
      {\pi}
}
%%%
\\
\localrule{ContextL}{
\lexec{\cloc, \lheap, \stack, e}{ \lheap', \stack', e'}{\pi}
}{
\lexec{\cloc, \lheap, \stack, E[e]}{\lheap', \stack', E[e']}{\pi}
}
%%%
\quad
\localrule{ContextR}{
  \lexec{\cloc_1, \lheap_1, \stack, e \parallel \cloc_2, \lheap_2}
        {\lheap_1', \stack', e' \parallel \lheap_2'}{\pi}
}{
  \lexec{\cloc_1, \lheap_1, \stack, E[e] \parallel \cloc_2, \lheap_2}
        {\lheap_1', \stack', E[e'] \parallel \lheap_2'}
        {\pi}
}
%%%
\end{gather*}
\caption{ Set of semantic rules described in 2.
  The left rules show the reduction of expressions that execute locally (a) and 
  the right rules, expressions that interact with remote objects (b). }
\label{fig:local}
\end{figure}

%% file: typing.tex
\label{sect:typing}

Figure~\ref{fig:typing} shows the typing rules of \lname.
They have the form $\typing{\Gammas}{e}{T}{\ctk}$ where 
an expression $e$ is verified against a sequence of typing 
contexts $\Gammas$ resulting in a type $T$ and an effect $\ctk$.
A typing context is a mapping from variables and addresses to types:
\begin{gather*}
\Gamma \in \entname{TypingContext} = 
(\entname{var}\ \cup\  \entname{Addr}) \rightarrow \entname{Type}
\end{gather*}
The effect $\ctk$ describes the behaviour of $e$, that is,
the memory accesses and messages sent to remote locations.
Effects are concatenated via the function $\pluseffect$ as defined below. 
\begin{gather*}
\Nothing \pluseffect \ctk = \ctk
\qquad
(\bop . \ctk_1) \pluseffect \ctk_2 = \bop . (\ctk_1 \pluseffect \ctk_2)
\qquad
\parallelT{\ctk_1}{\ctk_2} \pluseffect \ctk_3 = 
	\parallelT{\ctk_1 \pluseffect \ctk_3}{\ctk_2}
\end{gather*}
\input{typing_fig}
%
%The result of each rule is the type $T$ of the expression $e$ and
%the effect $\ctk$ (the behavioural type) 
%that 
%
The type $T$ associated to an expression is found in a standard way: 
similar can be found in~\cite{Clarke_oopsla12}, therefore we focus only in 
the behaviour produced.
The rules for variables and values, $\rulename{T-Var/Addr}$,
$\rulename{T-True/False}$, $\rulename{T-Skip/Null}$ result in 
empty effects, $\Nothing$, because they do not represent any
communication.
The typing rule $\rulename{T-Let}$ results in the
concatenation of the behaviour of both expressions.
%
%The effect produced in the return rule is the effect of typing the 
%returned expression.
%
The resulting behaviour of the rule $\rulename{T-Cond}$ is the 
behaviour of the predicate concatenated with a choice type 
which describes the behaviour of both branches.
The rule $\rulename{T-For}$ returns a loop type $\loopT{n}{\ctk}$, 
where $n$ is the number of iterations of the loop and $\ctk$ is
the behavioural type of its body.

The behaviour of the creation of an object, with $\rulename{T-NewO}$, 
is a $\keyword{write}$ behaviour, from the location of $\thisk$ to
the location of the new object, as new data is written to memory.
The predicate $\isActive{C}$ is true if the class of the object being 
created in annotated as active and the predicate $\isMain{\Gammas}{\thisk}$
is true if the class being verified is the main class. 
The field write is also represented by the $\keyword{write}$ behaviour,
given that it changes data already in memory. 
Typing the expression $\fwritec{e}{f}{e'}$ with the rule $\rulename{T-FWrite}$
returns the concatenation of the behaviour of $e$, the behaviour
of $e'$ and the $\keyword{write}$ from the location of $\thisk$
to the location of the object changed.
Following the same idea, the field read, $\freadc{e}{f}$, is represented 
by the $\keyword{read}$ behaviour and therefore the rule $\rulename{T-FRead}$
gives the  concatenation of the behaviour of $e$ with a $\keyword{read}$ type
from the location of $\thisk$ and to the location of the object read.
The typing rule, $\rulename{T-Call}$, describes synchronous method 
invocation which is only allowed if the receiver is in the same location
as the $\thisk$ object.
Its behaviour is the behaviour of the receiver concatenated with 
the behaviour of the expression passed as argument and the 
behavioural type annotated in the body of the invoked method.
The typing rule for the message send, $\rulename{T-Message}$, is similar.
However, it is possible to send a message 
to a different location and moreover it introduces parallelism in our types:
the receiving of the message should be executed in parallel with the 
continuation of the message sending---the resulting behaviour has the continuation
type, which in this case is $\Nothing$, in parallel with 
the expression to be executed due the message received.

%In order to simplify our formalisation we assume that mutually 
%recursive invocations (synchronous and asynchronous method calls)
%are not allowed. 
%The concatenation function used in the typing rules, 
%$\circ$, is defined below:

%% file: typing_fig.tex
\begin{figure}[!ht]
% \small
\begin{gather*}
\typerule{Var/Addr}
{}{
  \typing{\Gammas.\Gamma}{\vars}{\Gamma(\vars)}{\Nothing}\\
  \typing{\Gammas.\Gamma}{\Aaddr}{\Gamma(\Aaddr)}{\Nothing}
}
\quad
\typerule{True/False}{}{
  \typing{\Gammas}{\truek}{\boolk}{\Nothing}\\
  \typing{\Gammas}{\falsek}{\boolk}{\Nothing}
}
\quad
\typerule{Skip/Null}{\ }{
  \typing{\Gammas}{\skipk}{\nilk}{\Nothing}\\
  \typing{\Gammas}{\nullk}{\nilk}{\Nothing}
}
\quad
\typerule{Let}{ 
  \typing{\Gammas . \Gamma}
         {e_1}
         {T_1}
         {\ctk_1}
  \quad
  x \notin \dom{\Gamma}
  \\
  \typing{\Gammas. \Gamma[x \mapsto T_1]}
         {e_2}
         {T_2}
         {\ctk_2}
}{
\typing{\Gammas . \Gamma} {\letc{x}{e_1}{e_2}} {T_2} { \ctk_1 \pluseffect \ctk_2 }
}
%%%%%%%%%%%%%%%%%%%%%%
\\
\typerule{Cond}{
  \typing{\Gammas}{e_1}{\boolk}{\ctk_1}
  \quad
  \typing{\Gammas}{e_2}{T}{\ctk_2}
  \quad
  \typing{\Gammas}{e_3}{T}{\ctk_3}
}{
\typing{\Gammas}
       {\itec{e_1}{e_2}{e_3}}
       {T}
       {\ctk_1 \pluseffect \choiceT{\ctk_2}{\ctk_3}}
}
\quad
\typerule{For}{
  k > j
  \quad
  \Gammas = \Gammas'.\Gamma
  \quad
  \typing{\Gammas'.\Gamma[i \mapsto \intk]}
         {e}{T}{\ctk}
}{
\typing{\Gammas}
       {\forc{i}{j .. k}{e}}
       {T}{\loopT{k - j + 1}{\ctk}}
}
%%%
\\
\typerule{Ret}{
  \typing{\Gammas}{e}{T}{\ctk}
}{
  \typing{\Gammas . \Gamma}{\retk\ e}{T}{\ctk}
}
\quad
\typerule{NewO}{
  \isActive{C} \implies \isMain{\Gammas}{\thisk} 
  \quad
  \ots = \ownerT{C}{l_1, ..., l_n}
  \quad
  l_1 \neq ... \neq l_n
}{
  \typing{\Gammas}
         {\newk\ \ots}
         {\ots}
         { \writeT{\ell(\Gammas)}{l_1} }\\
}
%%%
\\
\typerule{FWrite}{
  \begin{array}[b]{c}
    \typing{\Gammas}{e}{\ownerT{C}{\sequence{l}}}{\ctk_1}
    \quad
    \replace{\flookup{C}{f}}{C}{\sequence{l}}  = T
    \quad
    \typing{\Gammas}{e'}{T}{\ctk_2}
  \end{array}
}{
\typing{\Gammas}{\fwritec{e}{f}{e'}}{T}
       {\ctk_1 \pluseffect \ctk_2 \pluseffect \writeT{\ell(\Gammas)}{l_1} }
}
\quad
\typerule{FRead}{
  \begin{array}[b]{c}
    \typing{\Gammas}{e}{\ownerT{C}{\sequence{l}}}{\ctk_1}
    \quad
    \replace{\flookup{C}{f}}{C}{\sequence{l}} = T
  \end{array}
}{
\typing{\Gammas}{\freadc{e}{f}}{T}
       {\ctk_1 \pluseffect \readT{\ell(\Gammas)}{l_1} }
}
\\
\typerule{Call}{
    \typing{\Gammas}{e_1}{\ownerT{C}{\sequence{l}}}{\ctk_1}
    \quad
    \typing{\Gammas}{e_2}{T'}{\ctk_2}
    \\
    \ell(\Gammas) = l_1
    \quad
    \replace{\mlookup{C}{m}}{C}{\sequence{l}} = (T, T', e_3, \ctk_3)
}{
\typing{\Gammas} %. \Gamma} 
       {\scallc{e_1}{m}{e_2}} 
       {T} 
       {\ctk_1 \pluseffect \ctk_2 \pluseffect \ctk_3}
}
\quad
\typerule{Message}{
  \begin{array}[b]{c}
    \typing{\Gammas}{e_1}{ \ownerT{C}{\sequence{l}} }{\ctk_1}
    \quad
    \typing{\Gammas}{e_2}{T'}{\ctk_2}
    \\
    \ell(\Gammas) = l_0
    \quad
    \replace{\mlookup{C}{m}}{C}{\sequence{l}} = (\nilk, T', e_3, \ctk)
  \end{array}
}{
\typing{\Gammas}{\acallc{e_1}{m}{e_2}}{\nilk}
       {\ctk_1 \pluseffect \ctk_2 \pluseffect \messageT{l_0}{l_1}{m}.\parallelT{\Nothing}{\ctk}}
}
\end{gather*}
\caption{Typing rules}
\label{fig:typing}
\end{figure}

%% file: main_results.tex
\label{sect:results}
We define a global behaviour, $\Sigma$, as a sequence
of behavioural types
\begin{gather*}\Sigma \in \sequence{\entname{Behaviour}}\end{gather*}
The behaviour of a node $\nodeconf$ describes the remote reads, writes and 
message sends to be executed by the node; it is obtained from the behaviour 
of the execution threads and message queues of all actors in that node.
%(cf. $\getSigmaAux{\nodeconf}{\sequence{\ctk}}$ from 
%Definition~\ref{def:global_behaviour}.
%$\getSigmaAux{\nodeconf}{\sequence{\ctk}}$ from Def. 4 in App. B).
%
The global behaviour of a runtime configuration, $\sequence\nodeconf$, describes 
the remote reads, writes and message sends to be executed by all nodes; it is the 
parallel combination of the behaviours of each the nodes $\nodeconf_i$. 
%(cf. $\getSigma{\sequence\nodeconf}{\Sigma}$ from 
%Definition~\ref{def:global_behaviour}).
Both definitions, the behaviour of a node and the global behaviour 
of a configuration, are below.
\input{get_global_behaviour}
%This context gives a global behaviour, the reads, writes and
%sends among different nodes.

Using this notion of global behaviour, 
we implicitly assume a well-formed program and we state soundness of 
our typing, which says that if a well-formed configuration,
$\sequence\nodeconf$, with a global behaviour $\Sigma$, 
reduces to another configuration $\sequence\nodeconf'$ through 
a communication step $\pi$ then the resulting configuration 
$\sequence\nodeconf'$ will have behaviour $\Sigma'$ which is a reduction
of $\Sigma$ through $\pi$.
\begin{theorem}
If $\wellFormedConfig{\sequence\nodeconf}
 \ands 
 \getSigma{\sequence\nodeconf}{\Sigma}  
 \ands 
 \gexec{\sequence\nodeconf}{\sequence\nodeconf'}{\pi}$
then  $\exists \Sigma': \getSigma{\sequence\nodeconf'}{\Sigma'} 
 \ands 
 \SigmaRelation{\Sigma}{\pi}{\Sigma'}$
\end{theorem}
% %%%%%%%%%%%%%%%%%%%%%%%%%%%
% %%%%%%%%%%%%%%%%%%%%%%%%%%%

The definitions of well-formed configuration (including well-formed heap
and well-formed stack) and (global) behaviour reduction 
are defined below:
\input{well_formed_config}
\input{behaviour_reduction}

Theorem 1 is a corollary of Lemmas 1 and 2.
\begin{lemma}
%\begin{gather*}
If $
\wellFormedh{\sequence\nodeconf}{\lheap}
 \ands
\wellFormedS{\lheap}{\stack}
 \ands
\lexec{\cloc, \lheap, \stack, e}{\lheap', \stack', e'}{\pi}
  \ands
\typingNoC{\lheap}{\stack}{e}{T}{\ctk}
  \ands
\neg (\stack\getInPos{2} = \Empty \ands e = \nullk) $
then $\exists \ctk': 
\typingNoC{\lheap'}{\stack'}{e'}{T}{\ctk'}
  \ands
\effectRelation{\filterf\ctk}{\filterf{\ctk'}}{\pi}$
%\end{gather*}
\end{lemma}

%%%%%%%%%%%%%%%%%
\begin{lemma}
%\begin{gather*}
If $
\wellFormedh{\sequence\nodeconf}{\lheap_1}
 \ands
\wellFormedh{\sequence\nodeconf}{\lheap_2}
 \ands
\wellFormedS{\lheap_1 \cup \lheap_2}{\stack}
 \ands
\lexec{\cloc_1, \lheap_1, \stack, e \parallel \cloc_2, \lheap_2}{\lheap_1', \stack', e' \parallel \lheap_2'}{\pi}
  \ands
\typingNoC{\lheap_1 \cup \lheap_2}{\stack}{e}{T}{\ctk}$
 then $\exists \ctk': 
\typingNoC{\lheap'_1 \cup \lheap'_2}{\stack'}{e'}{T}{\ctk'}
  \ands
\effectRelation{\filterf\ctk}{\filterf{\ctk'}}{\pi}$
%\end{gather*}
\end{lemma}

%% file: get_global_behaviour.tex
\begin{mydef}[The global behaviour]
\label{def:global_behaviour}
% \small
\begin{align*}
\textbf{(1) } &
\getSigma{\nodeconf_1, \dots, \nodeconf_n}{\sequence\ctk_1, \dots, \sequence\ctk_n} 
\textbf{ iff } 
  \forall i \in {1..n}: \getSigmaAux{\nodeconf_i}{\sequence\ctk_i} 
\\
%%%%%
\textbf{(2) } &
\getSigmaAux {\cloc, \lheap, \thread{\stack_1}{e_1}, \dots, \thread{\stack_n}{e_n}}
                            {\ctk_1 , \dots , \ctk_n} 
\textbf{ iff }
\forall i \in {1..n}: \getSigmaAux{\lheap,\stack_i,e_i}{\ctk_i}
\\
%%%%%
\textbf{(3) } &
\getSigmaAux{\lheap,\stack,e}{\filterf{\ctk \pluseffect \ctk_1 \pluseffect ... \pluseffect \ctk_n}} 
\textbf{ iff }
\exists T: \typingNoC{\lheap}{\stack}{e}{T}{\ctk} \ands 
(\lheap(\stack\getInPos{1}) = (C, \cloc^+, \_, m_1(v_1) \concat ... \concat m_n(v_n) \concat \Empty)
\\& \qquad  \ands 
\forall j \in {1..n}:
\exists T:
\typingNoC{\lheap}
          {(\thisk\mapsto\stack\getInPos{1}, x\mapsto v_j)}
          {\replace{\mlookup{C}{m_j}}{C}{\cloc^+}}
          {T}{\ctk_j})
\end{align*}
\end{mydef}

%% file: well_formed_config.tex
\begin{mydef}[Well-formed (1) configuration, (2) node, (3) heap, (4) stack 
and (5) stack frame]
\begin{align*}
\textbf{(1) }&
\wellFormedConfig{\sequence\nodeconf} 
\textbf{ iff } 
\forall i,j: 
  \nodeconf_i\getInPos{1} = \nodeconf_j\getInPos{1}
  \implies
  i = j
\ands
\forall \nodeconf': 
  \wellFormedNode{\sequence\nodeconf}{\nodeconf'}
%%%%%%%%%%%%%%%%%%%%%%%%%%%%%%%%%%%%%%%%%%%%%%%%%%%%
\\
\textbf{(2) }&
\wellFormedNode{\sequence\nodeconf}{\cloc, \lheap, ( \thread{\stack_1}{e_1}, ..., \thread{\stack_n}{e_n} )}
\textbf{ iff } 
\\ &\forall \alpha \in \dom{\lheap}:
  \alpha\getInPos{1} = \cloc 
  \ands
  \lheap(\alpha)\getInPos{2} = \cloc, \_
  \ands 
  \wellFormedh{\sequence\nodeconf}{\lheap} 
\\ & \ands
  \forall i \in \{1..n\}: 
  \wellFormedS{\getHeaps{\sequence\nodeconf}}{\stack_i} 
  \ands 
  \exists T_i, \ctk_i: \typingNoC{\lheap}{\stack_i}{e_i}{T_i}{\ctk_i}
\\
%%%%%%%%%%%%%%%%%%%%%%%%%%%%%%%%%%%%%%%%%%%%%%%%%%%%
\textbf{(3) }&
\wellFormedh{\sequence\nodeconf }{\lheap}
\textbf{ iff }
\forall \Aaddr \in \dom{\lheap}: 
  \wfValue{\getHeaps{\sequence\nodeconf } }
          {\Aaddr}
          {\getownership{\lheap}{\Aaddr}}
%%%%%%%%%%%%%%%%%%%%%%%%%%%%%%%%%%%%%%%%%%%%%%%%%%%%
\\
\textbf{(4) }&
\wellFormedS{\lheap}{\alpha \cdot \sframe_1, ..., \sframe_n}
\textbf{ iff }
\forall i \in \{1..n\}: \wellFormedS{\lheap}{\sframe_i}
\\
\textbf{(5) }&
\wellFormedS{\lheap}{(\thisk \mapsto \alpha, x_1 \mapsto v_1, \dots, x_n \mapsto v_n)}
\textbf{ iff }
\{\alpha, v_1...v_n\} \subseteq \{\truek, \falsek, \nullk \} \cup \dom{\lheap}
\end{align*}
\end{mydef}

%% file: behaviour_reduction.tex
\begin{mydef}[Global behaviour reduction]
% \small
\begin{align*}
\SigmaRelation{\Sigma}{\pi}{\Sigma'}
  \textbf{ iff } &
\Sigma = \sequence\ctk_1 , \ctk , \sequence\ctk_2 
  \ands
\Sigma' = \sequence\ctk'_1 , \ctk' , \sequence\ctk'_2 
  \ands
\effectRelation{\ctk}{\ctk'}{\pi}
  \ands \\ &
(\ctk = \parallelT{\ctk_1}{\ctk_2} 
  \implies
\ctk' = \ctk_1
  \ands
\exists \ctk_j \in \Sigma, \ctk'_j \in \Sigma' :
\ctk'_j = \ctk_j \pluseffect \ctk_2) 
\end{align*}
\end{mydef}

%%%%%%%%%%%%%%%%%%%%%%%%%%%%%%%%%%%%%%
%%%%%%%%%%%%%%%%%%%%%%%%%%%%%%%%%%%%%%

\begin{mydef}[Behaviour reduction] 
% \small
\begin{align*}
\effectRelation{\ctk_1}{\ctk_2}{\pi}
  \textbf{ iff } &
\ctk_1 = \pi . \ctk_2 \\
%%%
\effectRelation{\ctk_1}{\ctk_2}{\Nothing}
  \textbf{ iff } & 
\ctk_1 = \ctk_2 
  \ors 
\ctk_1 = \choiceT{\ctk_2}{\_} 
  \ors 
\ctk_1 = \choiceT{\_}{\ctk_2}
  \ors \\ &
(\ctk_1 = \loopT{n}{\ctk}. \ctk' \ands \ctk_2 = \ctk . \loopT{n-1}{\ctk} . \ctk') 
  \ors
\ctk_1 = \parallelT{\ctk_2}{\_}
\end{align*}
\end{mydef}

%% file: conclusion.tex
\label{sect:conclusion}

\paragraph{Related Work. }
To the best of our knowledge there is no integration of
behavioural types in the active/passive object paradigm,
or any formalism that combines behavioural types with ownership 
types to describe memory accesses;
however there are already a few programming 
languages that use session (behavioural) types in 
actor-based languages, namely: the integration of
session types in a Featherweight Erlang introduced by
Mostrous and Vasconcelos~\cite{mostrous.vasconcelos:session-typing-erlang};
an implementation of multiparty session types in an
actor library written in Python presented by Neykova and 
Yoshida~\cite{NY2014}; and the behavioural type system 
for an actor calulus, proposed by Crafa~\cite{crafa_actors}.

% there is no
%the closest work that we know uses session types in a
%compilation framework for distributed memory chip-level
%multiprocessing systems and was presented by Yoshida et 
%al.~\cite{yoshida.vasconcelos.etal:session-based-compilation}.
%
With respect to programming languages with the notion of locations
and proximity among processes and data, 
Rinard presented an extension of the programming language 
Jade (an implicitly parallel programming language designed 
to explore task-level concurrency~\cite{jade})
that allows the execution of tasks close 
to the data that they will use~\cite{locality_rinard}.
The language has constructs to describe how the processes access to 
the data; this information is analysed and used to improve the communication. 
Given that it is more expensive to access data remotely than locally, 
the author introduces a locality optimization algorithm that schedules 
the execution of tasks on places (processors) close to the data.
The programming language X10~\cite{Saraswat:x10}, developed by IBM, 
also features a notion of locality/places. 
In X10, each object can be either assigned to a place or distributed
among different places.
Chandra et al.~\cite{x10_locality:chandra} presented a new 
dependent type system for X10 that captures information 
about the locality of the resources in a partitioned heap
for distributed data structures, called \emph{place 
types}.
It provides information not only about whether a reference 
is local or remote, but also if two remote references point 
to resources in the same place or not.
Therefore, the compiler may use this information
to decrease the runtime overhead. 

\paragraph{Conclusion. }
This paper presents the fomalisation of a small oject-oriented
programming language that amalgamates behavioural types with
ownership types in order to describe remote memory 
accesses in NUMA systems.
Ownership types play a role in the representation of the 
topology and behavioural types in the definition of reads, 
writes and messages sent to remote locations. 
This sequence of memory access operations are annotated in
the method declarations as the ownership/location
parameters are annotated in class declarations.
This formalisation is just the first step towards a 
programming language that optimises performance by
moving objects to nodes where they have a cheaper cost
(the cost of interacting with other objects and of doing
remote accesses).

%% file: appendix.tex
%\section{Identifier Conventions}

\section{Identifier Conventions and Semantics}
\input{conventions}
%\input{only_local_semantics}
%\input{global_semantics}

\section{Auxiliary definitions, shorthands and lookup functions}
\input{definitions}

\input{value_agree}
\input{auxiliary}

\section{Topology Example}
\input{example}

%% file: conventions.tex
%\begin{figure}[!ht]
\paragraph{Identifier conventions. }
\begin{gather*}
n \in \natnumber 
\quad C \in \entname{ClassId}
\quad m \in \entname{MethId} 
\quad f \in \entname{FieldId} 
\quad L \in \entname{LocId}
\quad p \in \entname{OwnershipId}
\quad x,i \in \entname{varId}
\end{gather*}
%\caption{Identifier conventions}
%\end{figure}

%% file: definitions.tex
\begin{mydef}[Well-formed program and class]
\begin{align*}
\vdash P \equiv\ &
\forall ([\activek]\ \classk\ C \langle ... \rangle ... \in P): \wellFormedC{C}
& \quad
\wellFormedC{C}
 \equiv\ & 
\begin{cases}
\owners{C} = \{p_1, ..., p_n\}
 \ands \\
\forall m:
\mlookup{C}{m} = (T, x:T', e, \ctk) 
\ands \\
  \typing{(\thisk \mapsto \ownerT{C}{p_1, ..., p_n}, x \mapsto T')}
         {e}
         {T}
         {\ctk'}
\\ \implies
\ctk = \filterf{\ctk'}
\end{cases}
\end{align*}
\end{mydef}
Given that the effects returned during type checking do not exclude
reads and writes happening in the same node, we apply a function 
$\filterf{\ctk}$ in order to exclude such annotations.
The function is define as follows.
\small
\begin{gather*}
\filterf{\Nothing} = \Nothing
\qquad
\filterf{\parallelT{\ctk_1}{\ctk_2}} = \parallelT{\filterf{\ctk_1}}{\filterf{\ctk_2}}
\\
\filterf{\pi . \ctk} =
  (\text{if } \sourcef{\pi} = \destf{\pi} \text{ then } \Nothing \text{ else } \pi) . \filterf\commk
\\
\filterf{\choiceT{\commk_1}{\commk_2}.\commk_3} = 
  (\text{if } \filterf{\commk_1} = \Nothing \ands \filterf{\commk_2} = \Nothing 
    \text{ then } \Nothing 
    \text{ else } \choiceT{\filterf{\commk_1}}{\filterf{\commk_2}} ) . \filterf{\commk_3}
\\
\filterf{\loopT{n}{\commk} . \commk'} = 
  ( \text{if } \filterf\commk = \Nothing 
    \text{ then } \Nothing 
    \text{ else } \loopT{n}{\filterf\commk} ). \filterf{\commk'}
\end{gather*}
Note that if the expressiosns nested in for-loops or conditional
expressions have behaviour $\Nothing$, then the the loop or choice 
types are not annotated.
%%%%%%%%%%%%%%%%%%%%%%%%
%%%%%%%%%%%%%%%%%%%%%%%%

%% file: value_agree.tex
\begin{mydef}[Value agreement]
% \small
\begin{gather*}
\infer{WFTrue}{}{
\wfValue{\lheap}{\truek}{\boolk}
}
\quad
\infer{WFFalse}{}{
\wfValue{\lheap}{\falsek}{\boolk}
}
\quad
\infer{WFNull}{
T = \nilk \ors \isvalid{T}
}{
\wfValue{\lheap}{\nullk}{T}
}
\quad
\infer{WFObj}{
  \lheap(\Aaddr) = (C, 
      (\sequence{\cloc}),
      (f_i \mapsto v_i)_{ i \in I}, 
      \bullet)
  \\
  \forall i \in I : 
    \wfValue{\lheap}
            {v_i}
            {\replace{\flookup{C}{f_i}}{C}{\sequence{\cloc}}}
}{
\wfValue{\lheap}{\Aaddr}{\ownerT{C}{\sequence{\cloc}}}
}
\\
\infer{WFAObj}{
  \text{For }I\text{ some index set}
  \quad
  \lheap(\Aaddr) = (C, 
		  (\sequence{\cloc}),
		  (f_i \mapsto v_i)_{ i \in I}, 
		  m_1(v_1) \concat ... \concat m_n(v_n) \concat \Empty)
  \\
  \forall i \in I : 
    \wfValue{\lheap}
            {v_i}
            {\replace{\flookup{C}{f_i}}{C}{\sequence{\cloc}}}
   \quad
   %\forall \effect: 
   \typingNoC{\lheap}
              {\alpha \cdot (\thisk \mapsto \alpha, x \mapsto v_i)}
              {v_i}
              {\replace{\mlookup{C}{m_i}\getInPos{2}}{C}{\sequence\cloc}}{\ctk}
}{
\wfValue{\lheap}{\Aaddr}{\ownerT{C}{\sequence{\cloc}}}
}
%%
% \quad
% \infer{WFArray}{
%   \lheap(\Aaddr) = (\cloc_1, m, (i \mapsto v_i)_{i \in \{0.. m-1\}})
%   \\
%   \forall i \in \{0..m-1\}:
%     \wfValue{\lheap}
%             {v_i}
%             {\ownerT{C}{\cloc_1, \sequence\cloc}}
% }{
% \wfValue{\lheap}{\Aaddr}{\ownerT{C}{\cloc_1, \sequence\cloc}[m]}
% }
%%
\end{gather*}
\end{mydef}

%% file: auxiliary.tex
\paragraph{Lookup functions}

Considering $P$, the globally accessible program, and the class declaration
\begin{gather*}
\classk\ C \langle p^+\rangle \{ \sequence\fds\ \sequence\mds \} \in P
\end{gather*}
we define:
\begin{gather*}
\owners{C} = \{ p^+ \}
\\
\flookup{C}{f} = T 
  \textbf{ iff }
f: T \in \sequence\fds 
\\
\getFieldTypes{C} = \{ \sequence\fds \}
\\
\mlookup{C}{m} = (T, T', e, \ctk) 
 \textbf{ iff }
\defk\ m(x: T'): T\ \ink\ \ctk\ \{ e \} \in \sequence\mds
\\
\replace{\flookup{C}{f}}{}{l_1, ..., l_n} = 
\flookup{C}{f}[l_1/p_1, \dots, l_n/p_n ]
\textbf{ where } \owners{C} = \{p_1, \dots, p_n\}
\end{gather*}

\paragraph{Operations on the heap}
\begin{align*}
\lheap [ \Aaddr \mapsto o ] = \lheap' 
  & \textbf{ where } \lheap'(\Aaddr) = o \ands 
\forall \Aaddr_i \in \dom{\lheap}\setminus\{\Aaddr\}:\ \lheap(\Aaddr_i) = \lheap'(\Aaddr_i) 
\\
\lheap [\Aaddr, f \mapsto v  ] = \lheap' 
  & \textbf{ where } 
\lheap'(\Aaddr) = \lheap(\Aaddr)[f \mapsto v] \ands 
\forall \Aaddr_i \in \dom{\lheap}\setminus\{\Aaddr\}:\ \lheap(\Aaddr_i) = \lheap'(\Aaddr_i)
\\
%\lheap [\Aaddr, n \mapsto v  ] = \lheap' 
% & \textbf{ where } 
%\lheap'(\Aaddr) = \lheap(\Aaddr)[n \mapsto v]
%\\
\update{\lheap}{\Aaddr}{m(v)} = \lheap'
 & \textbf{ where }
\lheap (\Aaddr)= o 
 \ands 
o\getInPos{4} \neq \bullet
 \ands
\lheap' = \lheap [ \Aaddr \mapsto (o\getInPos{1}, o\getInPos{2}, o\getInPos{3}, m(v) \concat o\getInPos{4})]
\\
& \ands 
\forall \Aaddr_i \in \dom{\lheap}\setminus\{\Aaddr\}:\ \lheap(\Aaddr_i) = \lheap'(\Aaddr_i) 
\\
\getownership{\lheap}{\alpha} = \ownerT{C}{\sequence\cloc} 
& \textbf{ where }
\lheap(\alpha)\getInPos{1} = C
 \ands
\lheap(\alpha)\getInPos{2} = \sequence\cloc
\\
\lheap_1 \cup \lheap_2 = \lheap
& \textbf { where }
\forall \alpha \in \dom{\lheap}: \lheap_1(\alpha) = \lheap(\alpha) \ors \lheap_2(\alpha) = \lheap(\alpha)
\end{align*}

\paragraph{Operations on objects}
\begin{align*}
o(f) & \equiv o\getInPos{3}(f) \\
%ar(n) \equiv ar\getInPos{3}(n)\\
o[f \mapsto v] & \equiv (o\getInPos{1}, o\getInPos{2}, (f \mapsto v, \sequence{f_i \mapsto v_i}), o\getInPos{4})
  \qquad \textbf{where } o\getInPos{3} = f \mapsto \_, \sequence{f_i \mapsto v_i}\\
%ar[n \mapsto v] & \equiv (ar\getInPos{1}, ar\getInPos{2}, (n \mapsto v, \sequence{n_i \mapsto v_i}))
%  \qquad \text{ where } o\getInPos{3} = n \mapsto \_, \sequence{n_i \mapsto v_i}\\
\initialOb{\ownerT{C}{L_1, ..., L_m}} & \equiv 
\begin{cases}
(C, \cloc_1, ..., \cloc_m, (f_i \mapsto \initf{T_i})_{i \in 1..n}, \Empty) \quad \isActive{C}\\
(C, \cloc_1, ..., \cloc_m, (f_i \mapsto \initf{T_i})_{i \in 1..n}, \bullet) \quad \text{otherwise}
\end{cases}
\\ & \textbf{ where } \getFieldTypes{C} = \{f_1\colon T_1, \dots, f_n\colon T_n\}
\text{ and } \forall j \in \{1..m\}: \cloc_j = \locsMap(L_j)
\end{align*}

\paragraph{Operations on types}
\begin{gather*}
\initf{T} \equiv \text{if } T = \boolk \text{ then } \falsek \text{ else } \nullk
\qquad
\ell(\Gammas) = l \textbf{ iff } 
  \Gammas = \_ . \Gamma 
\ands
  \Gamma(\thisk) = \ownerT{C}{l, \_}
\end{gather*}

%%%%%% OTHER DEFS
\paragraph{Other definitions}
\begin{align*}
e [C, \cloc_1, \dots, \cloc_n ] & = e[\cloc_1/p_1, \dots, \cloc_n/p_n] \text{ where } 
\owners{C} = \{p_1, \dots, p_n\} \\
\getHeaps{\nodeconf_1, \dots, \nodeconf_n} = \lheap_1 \cup \dots \cup \lheap_n
  & \textbf{ iff }
\forall i \in \{1..n\}: \nodeconf_i\getInPos{2} = \lheap_i \\
\typingNoC{\lheap}{\stack}{e}{T}{\ctk} 
& \textbf{ iff }
\typing{\buildGammas{\lheap}{\stack} }{e}{T}{\ctk}\\
\typeOf{\lheap}{v} &\equiv
\text{if } v=\truek \ors v=\falsek \text{ then } \boolk \text{ else } \getownership{\lheap}{v}
\\
\inferNN{
  \buildGammas{\lheap}{ \sframe_1 } = \Gamma_n
  \\ \dots \\
  \buildGammas{\lheap}{ \sframe_n } = \Gamma_1
}{
  \buildGammas{\lheap}{ \alpha \cdot \sframe_1 \dots \sframe_n }
}
& \quad
\inferNN{
  T_{this} = \typeOf{\lheap}{\alpha}
  \\
  T_1 = \typeOf{\lheap}{v_1}
  \quad
  \dots
  \quad
  T_n = \typeOf{\lheap}{v_n}
  \\
  \Gamma = (\thisk \mapsto T_{this}, x_1 \mapsto T_1, \dots, x_n \mapsto T_n)
}{
  \buildGammas{\lheap}
              { \thisk \mapsto \alpha, x_1 \mapsto v_1, \dots, x_n \mapsto v_n } 
= \Gamma
}
\end{align*}

%% file: example.tex
Consider the following code with three class declarations:
an active class \lstinline!C!, a passive \lstinline!D! and 
the class \lstinline!Main!.
An active object, instance of \lstinline!C!, has three fields
pointing to three objects in different locations of type 
\lstinline!D!.
The class main creates three abstract (or symbolic) locations
\lstinline!L1, L2, L3! and the body of the \lstinline!main! method.

\begin{minipage}{0.4\textwidth}
\begin{lstlisting}[numbers=none]
active class C<p1, p2, p3> 
	d1: D<p1>
	d2: D<p2>
	d3: D<p3>

class D<p>
\end{lstlisting}
\end{minipage}
\begin{minipage}{0.6\textwidth}
\begin{lstlisting}[numbers=none]
class Main<L1, L2, L3>
	def main(): nil 
		as b write(L1, L2). write(L1, L3) {
		let x = new C <L1, L2, L3> in 
		let y = (x.d1 = new D<L1>) in 
		let z = (x.d2 = new D<L2>) in 
                x.d3 = new D<L3>
	}
\end{lstlisting}
\end{minipage}

\begin{figure}[!ht]
\begin{center}
\includegraphics[scale=0.56]{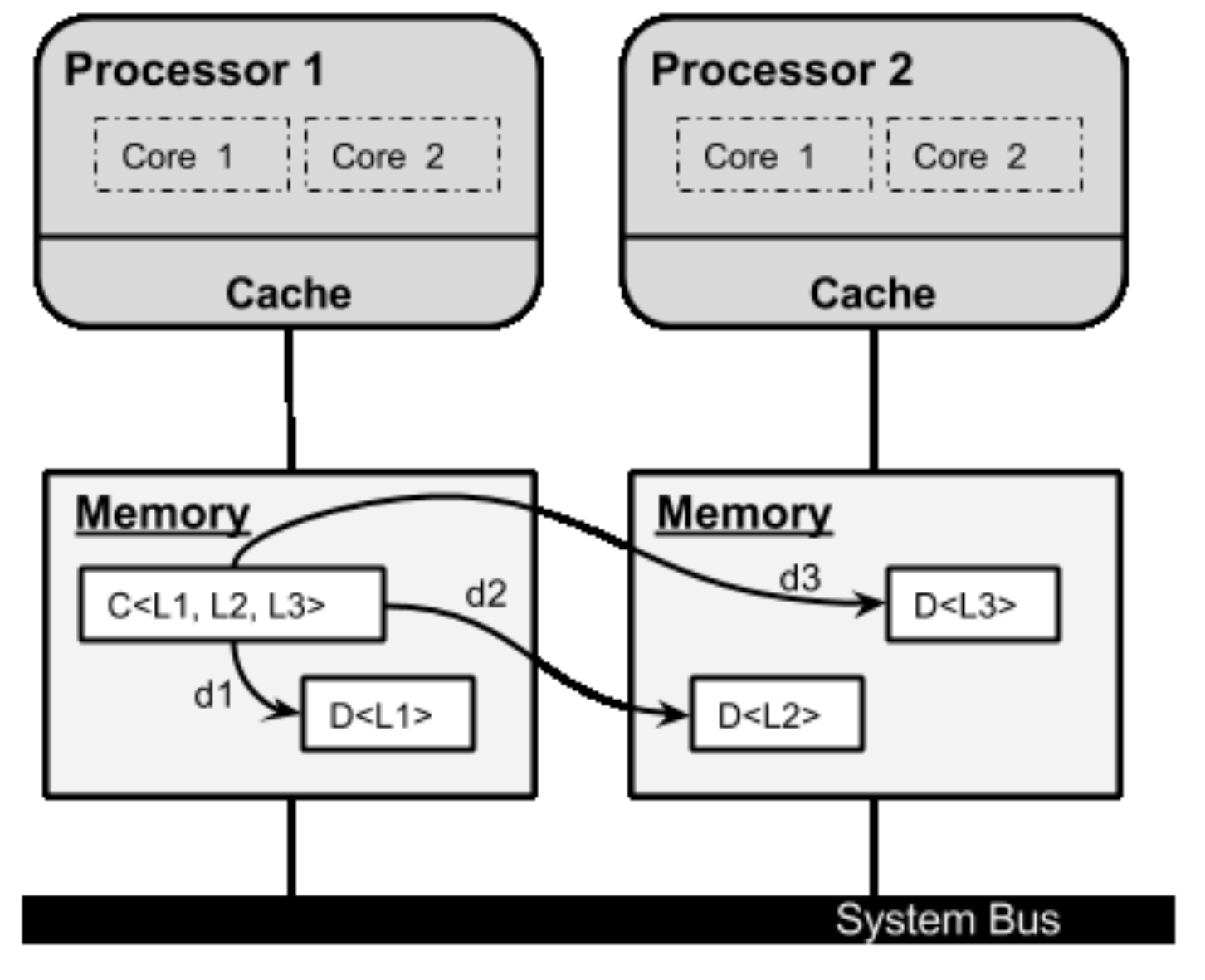}
\end{center}
\caption{NUMA system with two different nodes}
\label{fig:NUMA_memory}
\end{figure}

The topology after execution of the main method
is depicted in the following figure.
\begin{figure}[!ht]
\begin{center}
\includegraphics[scale=0.6]{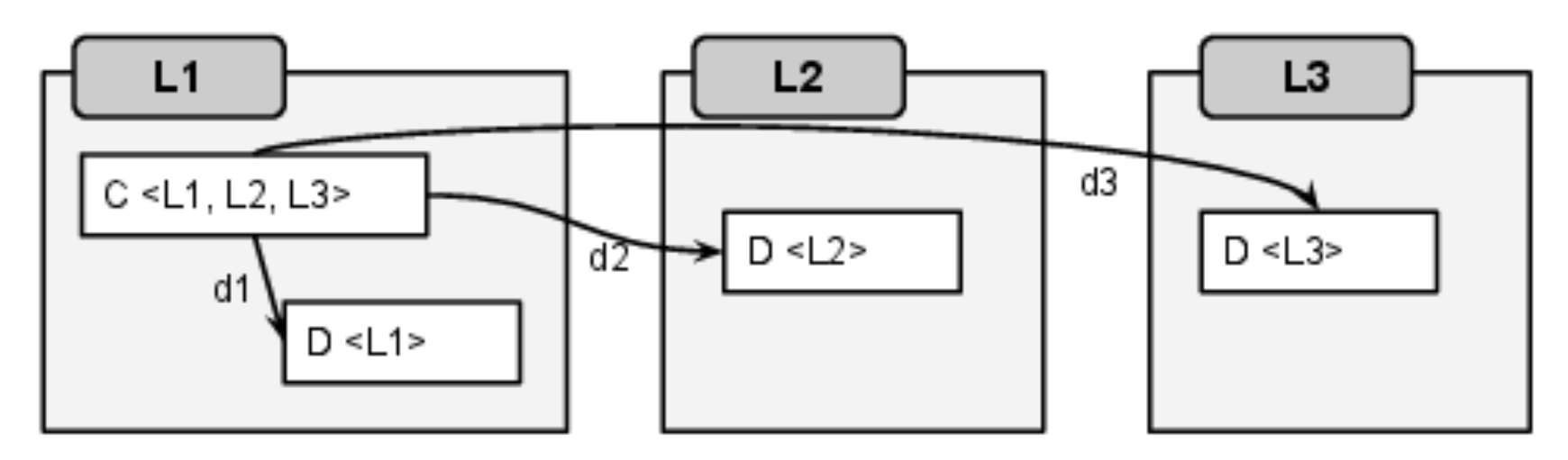}
\end{center}
\caption{The ownership topology after the execution of the
			expression in the method \lstinline!main!.}
\end{figure}
In the abstract location \lstinline!L1! there is an instance
of class \lstinline!C! and an instance of class
\lstinline!D!. Abstract locations \lstinline!L2! and
\lstinline!L3! have both an instance of class \lstinline!D!.
Although the programmer define 3 abstract locations, the machine
might have a different number of nodes.
For instance, in a system with two different nodes, 
we could have the mapping
$(L_1 \mapsto \cloc_1, L_2 \mapsto \cloc_2, L_3 \mapsto \cloc_2)$
between abstract locations and node identifiers, which means that the objects
in \lstinline!L1! are in the
node $\cloc_1$, and objects from \lstinline!L2! and \lstinline!L3!
are in the same node, as depicted in Figure~\ref{fig:NUMA_memory}.
